\documentclass[prd,twocolumn,showpacs,preprintnumbers]{revtex4}

\newcommand{\A}{A}
\newcommand{\B}{B}
\newcommand{\C}{C}
\newcommand{\D}{D}

\newcommand{\td}{{\tilde d}}

\newcommand{\e}{{\rm e}}
\newcommand{\de}{ \Delta}

\newcommand{\p}{{\partial}}

\newcommand{\lb}{\label}
\newcommand{\be}{\begin{equation}}
\newcommand{\ee}{\end{equation}}
\newcommand{\bea}{\begin{eqnarray}}
\newcommand{\eea}{\end{eqnarray}}
\newcommand{\bw}{\begin{widetext}}
\newcommand{\ew}{\end{widetext}}

\begin{document}
\preprint{DTP-MSU/04-19} \preprint{LAPTH-1080/04}
\preprint{hep-th/0412321}
\title{Black branes on the linear
dilaton background}
\author
{G\'erard Cl\'ement$^{a}$, Dmitri Gal'tsov$^{a,b}$ and C\'edric
Leygnac$^{a}$}\email{gclement@lapp.in2p3.fr; gdmv04@mail.ru;
leygnac@lapp.in2p3.fr
$^{a}$Laboratoire de  Physique Th\'eorique LAPTH (CNRS),
\\
B.P.110, F-74941 Annecy-le-Vieux cedex, France\\ $^{b}$Department
of Theoretical Physics, Moscow State University, 119899, Moscow,
Russia}
\begin{abstract}
We show that the complete static black $p$-brane supergravity
solution with a single charge  contains two and only two branches
with respect to behavior at infinity in the transverse space. One
branch is the standard family of asymptotically flat black branes,
and another is the family of black branes which asymptotically
approach the linear dilaton background with antisymmetric form flux
(LDB). Such configurations were previously obtained in the
near-horizon near-extreme limit of the dilatonic asymptotically flat
$p$-branes, and used to describe the thermal phase of field theories
involved in the DW/QFT dualities and the thermodynamics of little
string theory in the case of the NS5-brane. Here we show by direct
integration of the Einstein equations that the asymptotically LDB
$p$-branes are indeed exact supergravity solutions, and we prove a
new uniqueness theorem for static $p$-brane solutions satisfying
cosmic censorship. In the non-dilatonic case, our general
non-asymptotically flat $p$-branes are uncharged black branes on the
background $AdS_{p+2}\times S^{D-p-2}$ supported by the form flux. We
develop the general formalism of quasilocal quantities for
non-asymptotically flat supergravity solutions with antisymmetric
form fields, and show that our solutions satisfy the first law of
theormodynamics. We also suggest a constructive procedure to derive
rotating asymptotically LDB brane solutions.
\end{abstract}
\maketitle
\section{Introduction}
The existence of two alternative descriptions of branes, the
classical one within supergravity theories, and the quantum one in
string theory, leads to various holographic dualities between
classical supergravities and quantum field theories, from which
the AdS/CFT correspondence \cite{adscft} was originally
discovered. This correspondence relates to non-dilatonic branes,
such as the M2 and M5 branes of the M-theory and the D3 brane of
string theory, which have an AdS  near-horizon structure. The
asymptotic boundary of the AdS space is conformal to the Minkowski
space-time where the dual conformal field theory lives.

The AdS/CFT conjecture was extended later to the generic case of
string-theoretical dilatonic branes, in which case the
near-horizon geometry is either AdS or Minkowski  with a
non-trivial dilaton field  depending  linearly on an appropriate
radial coordinate. Such configurations are also supersymmetric in
the context of supergravities (though not maximally supersymmetric
as in the case of non-dilatonic branes), but the conformal
symmetry is broken by the dilaton. These backgrounds are dual to
non-conformal QFT-s with sixteen supercharges living on their
boundary \cite{ItMaSoYa98}. In the case of the NS5 brane
\cite{DuLu91,CaHaSt91}, the corresponding dual theory is not a
local field theory, but the so-called little string theory
\cite{Se97} (LST) living on the flat six-dimensional world-volume
of the $NS5$-brane in the string frame (for a review and recent
references see \cite{Ah99}).

More general considerations were presented in \cite{Be99}
(extending the previous work of \cite{BoSkTo99}) for any
dimensions and various fractions of supersymmetry. It was argued
that although the near-horizon geometry of the extremal dilatonic
brane is singular, by transforming it to the so-called "dual"
frame (the Nambu-Goto frame for the dual brane probe) one obtains
the product of  an AdS space with a sphere. After reduction over
the sphere one gets the domain wall (DW) solution, for which
reason the corresponding duality was termed as the DW/QFT
correspondence \cite{BoSkTo99}. The  near-horizon configuration of
the generic dilatonic brane is the product of either AdS, or flat
space-time with a sphere endowed with a non-trivial dilaton. In
what follows we will call this field configuration the linear
dilaton background (LDB) independently of any particular frame or
coordinate system used.

By the standard argument, the {\it thermal} version of the dual
quantum theory should have as a holographic dual the linear dilaton
background endowed with  an event horizon.  Such a configuration was
obtained for the NS5 case (dual to LST) by Maldacena and Strominger
in the near-horizon limit of the near-extremal NS5-brane
\cite{MaSt97} and for a discrete family of  rotating dilaton branes
by Harmark and Obers \cite{HaOb00}. A similar four-dimensional
``horizon-plus-throat'' geometry was presented earlier by Giddings
and Strominger \cite{GS} (see also \cite{KP}) as a certain limit near
the horizon of the near extremal dilaton black hole \cite{Gi82}.  The
relation between the linear dilaton background and the
horizon-plus-throat geometry is similar to the relation between the
anti-de-Sitter space and the Schwarzschild-anti-de Sitter black hole.
This configuration was shown to be a fully legitimate solution of the
Einstein-Maxwell-dilaton four-dimensional theory, thus extending the
family of asymptotically flat and asymptotically AdS/dS black holes
to the asymptotically LDB solutions \cite{ChHoMa94,ClGaLe02,ClLe04}.
In addition, in  Refs. \cite{ClGaLe02,ClLe04} various generalizations
(including rotation) of black holes with linear dilaton asymptotics
were obtained. Similar solutions exist in presence of a dilaton
potential \cite{ChHoMa94,CaWa04}.

The purpose of the present paper is to study systematically the
{\em higher-dimensional} asymptotically LDB solutions to the
supergravity theories. We  reexamine the generic supergravity
equations for the metric, dilaton and the antisymmetric form in
$D$ dimensions for arbitrary values of the form rank and the
dilaton coupling constant, assuming the $R\times ISO(p)$ symmetry
of the world-volume and the $SO(D-p-1)$ symmetry of the transverse
space and imposing no further restrictions on the ansatz. The
corresponding system of equations is fully integrable, and,
following  Ref. \cite{GaLeCl04}, we obtain the generic solution
containing a number of integration constants. Assuming the
existence  and regularity of the (non-degenerate) event horizon,
we reduce the number of free parameters to three and find that the
metric functions become constant at infinity unless some special
condition on the parameters is imposed. In this special case one
obtains the solutions with  linear dilaton asymptotics. In the
first generic case, after trivial coordinate rescalings, one
arrives at the usual asymptotically flat  $p$-branes
\cite{HoSt91,DuKhLu95,GaRy98}. Thus the asymptotically LDB
$p$-branes form a degenerate family of solutions in the full
solution space. Remarkably, no other alternatives regarding the
admissible asymptotic behavior exist for $p$-branes with regular
horizons within the theory with no dilaton potential and no
cosmological constant.

Our general family of non-asymptotically flat branes includes the
non-dilatonic case. In this case one deals with neutral black
$p$-branes on the  background $AdS_{p+2}\times S^{D-p-2}$ supported
by the flux of antisymmetric form. In particular, we obtain the
${\tilde {\rm M}}2$ and ${\tilde {\rm M}}5$ branes of M-theory
approaching at infinity  $AdS_4\times S^7$ and $AdS_7\times S^4$
respectively, and the ${\tilde {\rm D}}3$-brane of IIB theory
approaching $AdS_5\times S^5$. These ``tilde'' $p$-branes interpolate
between the product of  flat space and a sphere at the horizon, and
the product of the anti de Sitter space and a sphere at infinity. The
``tilde'' $p$-branes are not supersymmetric unless the mass parameter
is set zero, in which case we obtain the linear dilaton background
with a flux.

To calculate the brane tension and other physical characteristics
of the asymptotically non-flat solutions one needs to generalize
the formalism of quasilocal charges developed in particular in
Refs. \cite{BY,HaHo95,ChNe98} to the case of an arbitrary number
of space-time dimensions and to the presence of the antisymmetric
form fields. We perform this and apply it to the case of branes on
the linear dilaton background.

Summarizing our results, we formulate a generalized uniqueness
theorem claiming that static brane solutions without naked
singularities possessing the $R\times ISO(p)\times SO(D-p-1)$
isometries exist in two and only two realizations: one is the
usual family of  asymptotically flat $p$-branes and another is the
family of branes which asymptotically approach the linear dilaton
background. The first family has a BPS limit, while for the
second one the BPS limit coincides with the linear dilaton
background itself. We conjecture that the same should be true for
the intersecting branes as well.

\section{Linear dilaton background with flux}
We consider the  action  containing the graviton, a $q$-form field
strength, F$_{[\td+1]}$, and a dilaton, $\phi$, coupled to a form
field with the coupling constant $a$:
\begin{equation}\label{action}
S = \int d^D x \sqrt{-g} \left( R - \frac12 \partial_\mu \phi
\partial^\mu \phi - \frac{{\rm e}^{a\phi}}{2\, (\td+1)!} \,  \,
F_{[\td+1]}^2
\right)
\end{equation}
(throughout this paper, the Newton constant $G$ is set to the
value $1/16\pi$).
In view of the electric-magnetic duality of the corresponding
equations of motion
\begin{equation} \label{duality}
g_{\mu\nu} \to g_{\mu\nu}, \qquad F_{\td+1} \to {\rm e}^{a\phi} \ast
F_{D-\td-1}, \qquad \phi \to -\phi,
\end{equation}
we will restrict ourselves here to the magnetic solutions.

The asymptotically flat static BPS solution with  ISO($p+1$)
symmetry of the world volume and SO($\td+$2) symmetry of the
transverse space, $(D=p+\td+3)$, reads \cite{HoSt91,GaLeCl04} \bea
&&ds^2=
f^{\frac{4\td}{\Delta(D-2)}}ds^2_{p+1}+f^{\frac{2a^2}{\Delta\td}}
\left(f^{-2}dr^2+r^2d\Omega^2_{\td+1}\right),\nonumber\\
&&\e^{\phi-\phi_{\infty}}=f^{2a/\Delta},\nonumber\\
&&F_{[\td+1]}=\frac{2\td}{\sqrt{\Delta}}r_0^\td\e^{-a\phi_{\infty}/2}
\, {\rm vol}(\Omega_{\td+1}),\lb{afsol}\eea where \be
ds^2_{p+1}=-dt^2+d{\bf x}_p^2\ee is the world-volume flat
space-time interval, \be \lb{Delt} \Delta=a^2+\frac{2d\td}{D-2},
\ee ($d = p+1$) and \be f=1- \left(\frac{r_0}{r}\right)^\td .\ee
It depends on two real parameters $r_0,\,\phi_{\infty}$ --- the
horizon radius   and the asymptotic value of the dilaton. Shifting
the horizon from $r=r_0$ to $r=0$ via the coordinate
transformation \be r\to \left(r^\td+r_0^\td\right)^{1/\td}\ee one
obtains \bea &&ds^2=
H^{\frac{-4\td}{\Delta(D-2)}}ds^2_{p+1}+H^{\frac{4d}{\Delta(D-2)}}
\left( dr^2+r^2d\Omega^2_{\td+1}\right), \\
&&\e^{\phi-\phi_{\infty}}=H^{-2a/\Delta},\quad H=1+
\left(\frac{r_0}{r}\right)^\td. \lb{dil}\eea

The desired near-horizon limit of this solution can be  obtained
by omitting the constant in the harmonic function $H$:\bw\bea
\lb{eldb} ds^2=\left(\frac{
r}{r_0}\right)^{\frac{4\td^2}{\Delta(D-2)}}ds_{p+1}^2
+\left(\frac{r_0}{r}\right)^{\frac{4d\td}{\Delta(D-2)}}
\left( dr^2 +r^2d\Omega^2_{\td+1} \right), \nonumber\\
\e^{(\phi-\phi_{\infty})}=
\left(\frac{r}{r_0}\right)^\frac{2a\td}{\Delta} ,\quad
F_{[\td+1]}=\frac{2\td}{\sqrt{\Delta}}r_0^\td\e^{-a\phi_{\infty}/2}\,
{\rm vol}(\Omega_{\td+1}). \eea\ew This is the solution which we
will call LDB presented in the Einstein frame. For supergravity
theories, admitting the 1/2 BPS asymptotically flat $p$-brane
solutions (in which case  $\Delta=4$),  the linear dilaton
background  is 1/2 supersymmetric as well, unless $a=0$ in which
case  supersymmetry is fully restored in the near horizon limit.
This solution is supported both by the dilaton and the
antisymmetric form flux.

In the Einstein frame the metric (\ref{eldb}) does not have a
clear geometric meaning as $r\to\infty$. However, as was clarified
in \cite{DuGiTo94,Be99}, the space-time always has the AdS
structure in the so-called ``dual frame', which is the Nambu-Goto
frame for the dual brane probe. Defining the dual frame
\cite{Be99} by the conformal transformation \be \lb{todual}
ds^2_{\rm dual} =\e^{-a(\phi-\phi_{\infty})/\td}ds^2,\ee and
passing to a new radial coordinate \be \lambda=r_0\ln(r/r_0),\ee
we find \be \lb{dsdual} ds^2_{\rm
dual}=\e^{\frac{2(2\td-\Delta)\lambda}{r_0\Delta}}ds_{p+1}^2+d\lambda^2+r_0^2
d\Omega^2_{\td+1}.\ee In terms of the new radial coordinate the
dilaton is precisely linear: \be
\phi-\phi_{\infty}=\frac{2a\td}{\Delta r_0}\lambda.\ee In the
particular case \be\Delta=2\td,\quad \mbox{or}\quad
a^2=\frac{2\td^2}{D-2}, \ee the space-time (\ref{dsdual}) becomes
the direct product of the $p+2$ dimensional Minkowski space and a
sphere. In ten-dimensional supergravities (with $\Delta=4$) this
corresponds to fivebranes.

For $\Delta\neq 2\td$ the metric (\ref{dsdual}) is the product
$AdS_{p+2}\times S^{\td+1}$. Introducing the horospherical
coordinate \be u=\frac{\e^{\frac{q\lambda}{r_0}}}{r_0\,q},\ee
where \be q=\frac{\Delta}{2\td-\Delta},\ee it can be cast into the
more familiar form \be ds^2_{\rm dual}= r_0^2\left( (qu)^2
ds_{p+1}^2 +\frac{du^2}{(qu)^2}+ d\Omega^2_{\td+1}\right).
\lb{u}\ee

It should be stressed  that the LDB solution (\ref{eldb}) is
supported not only by the dilaton field, but by the field of the
antisymmetric form as well. An important particular case of the
above considerations is $a=0$, when there is no dilaton at all.
Then the dual frame coincides with the Einstein frame, so the
solution (\ref{u}) is an exact solution in the Einstein frame. In
fact, in $D=11$ supergravity this corresponds to $AdS_4\times S^7$
and $AdS_7\times S^4$, while in the D3 sector of type IIB theory
--- to $AdS_5\times S^5$. These are fully supersymmetric
solutions of the corresponding theories. Being supported by the
form flux, strictly speaking  they do not belong to the class of
linear dilaton backgrounds. We still call the solution
(\ref{eldb}) ``LDB'' for any values of the parameters $d,\,
\td,\,a$ simply because of its generality.
\section{Multicenter generalization}
The transverse part of the Einstein frame LDB metric (\ref{eldb})
is conformally flat, this suggests its multicenter generalization.
To put the derivation into a constructive form, we invoke the
sigma-model formulation of the problem \cite{GaRy98}.

Consider the class of metrics with the conformally flat world
volume part \be
ds^2=\e^{(\psi-a\phi)/d}ds^2_{p+1}+\e^{(a\phi-\psi)/\td}h_{ij}dx^i
dx^j,\ee where $ds^2_{p+1}$ is Minkowskian, and a new scalar
$\psi$ is introduced. It is also convenient to use as the second
independent scalar the linear combination \be
\xi=\frac{D-2}{2}\left(  a\psi -\Delta \phi \right)\ee instead of
the dilaton $\phi$.  The antisymmetric form in the magnetic case
can be parameterized by the scalar $u$ via \be F=\e^{-\psi}\star
du,\ee where Hodge dualization is understood with respect to the
transverse $\td+2$ dimensional space. All scalar quantities $\psi,
\xi, u$ are assumed to depend only on coordinates $x^i$,
parameterizing the transverse space.

Performing the  dimensional reduction we obtain the $\td+2$
dimensional gravitating sigma-model  \be \label{act_M}
S_{\sigma}=\int\,\left(R(h)+\frac{1}{\Delta}h^{ij}
\mbox{Tr}\left(\partial_i M\partial_j
M^{-1}\right)\right)\sqrt{h}d^{\td+2}x, \ee where the matrix $M$
\be \label{M} M=\e^{-\frac{\psi}{2}}\left(
\begin{array}{ccc}
2& \frac{u\sqrt{\Delta}}{2}&0\\
\frac{u\sqrt{\Delta}}{2}& \frac{u^2\Delta}{8}-\frac{\e^{\psi}}{2}&0\\
0&0&\e^{\frac{\psi +\nu\xi}{2}}\\
\end{array}
\right)\;\ee parametrizes the target space $SL(2,R)/SO(1,1)\times
R$. In this formula $\nu=\frac{2}{\sqrt{(D-2)d\td}}$. The
corresponding equations of motion read \bea \label{eq_M}
&\frac{1}{\sqrt{h}}\partial_i(\sqrt{h}h^{ij}M^{-1}
\partial_j M)=0,&\\
 \label{eq_R} &R_{ij}(h)=-\Delta^{-1}\mbox{Tr}(\partial_i M
\partial_j M^{-1}).& \eea This representation is a convenient
starting point for an application of the harmonic map technique.

It was noticed that the BPS solutions can be presented as the null
geodesics of the target space \cite{Clem,C_G}. If the matrix $M$
depends on $x$ through a single function, $M=M(\sigma(x))$, with
$\sigma(x)$ being a harmonic function on the curved space with the
metric $h$ \be \frac{1}{\sqrt{h}}\partial_i(\sqrt{h}h^{ij}
\partial_j \sigma)=0,
\ee the equation (\ref{eq_M}) reduces to the matrix equation \be
\frac{d}{d\sigma}\left(M^{-1}\frac{dM}{d\sigma}\right)=0, \ee
whose solution  can be expressed  in the exponential form \be
\label{exp_M} M=M_0{\rm e}^{K\sigma}, \ee where $K$ belongs to the
Lie algebra of the group $G$, and $M_0$ is a constant matrix
corresponding to the value of $M$ at some normalization point.

Substituting this into the Einstein equations (\ref{eq_R}) one
gets \be R_{ij}(h)=\Delta^{-1}\mbox{Tr}(K^2)\partial_i \sigma
\partial_j \sigma.
\ee It follows that in the particular case $\mbox{Tr}(K^2)$=0 the
metric $h$ is Ricci-flat. This is a constructive way to build
null-geodesic solutions to an arbitrary $\sigma$-model.

There are two distinct classes of solutions depending on whether
$\det K$ is zero or not. In the first (degenerate) case, taking\be
K=\left(\begin{array}{ccc}
 1&1/2 &0\\-2&-1&0\\ 0&0&1\\
\end{array}\right),\;\;M_0=\left(\begin{array}{ccc}
 2&0 &0\\0&-1/2&0\\ 0&0&1\\
\end{array}\right),\ee
one obtains \be M=\left(\begin{array}{ccc}
2(1+\sigma)&\sigma&0\\ \sigma&-\frac12  (1-\sigma)&0\\ 0&0&1\\
\end{array}\right),
\ee where $\sigma$ is a harmonic function in $\td+2$ dimensional
Ricci-flat space. Comparing this with the Eq. (\ref{M})   we get
\be \psi=-2\ln(1+\sigma),\quad \xi=0,\quad
u=\frac{2}{\sqrt{\Delta}} \frac{\sigma}{1+\sigma}, \ee which
corresponds to the  metric \be
ds^2=(1+\sigma)^{\frac{-4\td}{\Delta
(D-2)}}ds^2_{p+1}+(1+\sigma)^{\frac{4 d}{\Delta (D-2)}}h_{ij}dx^i
dx^j.\ee  This is the usual BPS $p$-brane solution with the
harmonic function $H=1+\sigma$ in the Ricci-flat transverse space.
The corresponding  dilaton field is given by ($\phi_\infty=0$)\be
{\rm e}^{a\phi}=(1+\sigma)^{-2a^2/\Delta}. \ee  The harmonic
function has the Coulomb form once the transverse space is chosen
flat $h_{ij}=\delta_{ij}$:\be
H=1+\sigma=1+\left(\frac{r_0}{r}\right)^\td.\ee The LDB solution
(\ref{eldb}) corresponds to replacing $H$ by its limit for $r \to
0$:\be H=1+\sigma=\left(\frac{r_0}{r}\right)^\td.\ee Clearly, it
admits the following multicenter generalization \be
1+\sigma=\sum_n \frac{c_n}{|{\bf r}-{\bf r}_n|^\td},\ee where
$c_n$ is a set of real constants. Alternatively, we can express
the multicenter LDB solution in the form \be
M=M'_0\e^{K\sigma'},\quad \mbox{with}\quad \sigma'=\sum_n
\frac{c_n}{|{\bf r}-{\bf r}_n|^\td},\ee i.e., $\sigma'=1+\sigma$,
leading to a non-diagonal $M_0'=M_0\e^{-K}$, such as used in
\cite{ClLe04}.

In the case of a non-degenerate generating matrix $ \mbox{det}\;
K\ne 0$ one gets solutions with naked singularities, or geodesically
complete solutions, for details
see Refs. \cite{GaRy98,ClLe04}.

\section{General supergravity solution}
Let us now  pass to a more general formulation of the problem. We
wish to study the $p$-brane solutions to the action (\ref{action})
whose world volume is given by the $d = p+1$ dimensional space
with the isometries $ISO(p)\times R$ and whose transverse space is
spherically symmetric. The line element
\begin{equation}\label{metric}
ds^2 = - {\rm e}^{2\B} dt^2 + {\rm e}^{2\D} d{\bf y}_p^2 + {\rm
e}^{2\C} \, d\Omega_{\td+1}^2 + {\rm e}^{2\A} d\rho^2,
\end{equation}
is parametrized by four  functions $\A(\rho),\,\B(\rho),\,
\C(\rho)$ and $\D(\rho)$. Assuming this ansatz, the equations for
the form field and the corresponding Bianchi identity \bea
&&\partial_\mu \left( \sqrt{-g} \, {\rm e}^{a\phi} \,
F^{\mu\nu_1\cdots\nu_\td} \right) = 0, \label{form}\\
&&\partial_\mu \left( \sqrt{-g} \,  \ast F^{\mu\nu_1\cdots\nu_d}
\right) = 0, \label{bianchi}\eea (where dualization is understood
with respect to the full D-dimensional space-time) can easily be
solved in the magnetic sector as \be
F_{a_1,...,a_{\td+1}}=b\,\sqrt{{\bar
g}}\,\epsilon_{a_1,...,a_{\td+1}}, \label{solF}\ee or in short
notation \be F_{[\td+1]} = b \,\, \mbox{vol}(\Omega_{\td +1 }),
\end{equation}
where $b$ is the constant field strength parameter  and $\bar
g_{ab}$ is the metric on the unit $\td+1$ dimensional sphere.

The system of equations was derived in the previous papers
\cite{ChGaGu02,GaLeCl04}, for convenience we present it here in
the current notation. The Ricci tensor for the metric
(\ref{metric}) has the following non-vanishing components \bw
\begin{eqnarray}
R_{tt} &=& {\rm e}^{2\B-2\A} \left[   \B'' + \B' (\B'-\A'+(\td
+1)\C'+(d-1)\D' )  \right], \label{Rtt}\\ R_{xx} &=&
\e^{2D-2A}\left[- \D'' - \D'(\B'-\A'+(\td +1)\C'+(d-1)
\D')\right],\\ R_{rr} &=& - \B'' - \B'(\B'-\A') - (\td
+1)(\C''+\C'^2-\A' \C') - (d-1)(\D''+\D'^2-\A' \D'), \lb{Rrr} \\
R_{ab} &=& \left\{- {\rm e}^{2\C-2\A} \left[  \C'' +
\C'(\B'-\A'+(\td +1)\C'+(d-1)\D' ) \right] +  \td \right\} \, \bar
g_{ab}, \label{Rab}
\end{eqnarray}\ew
where primes denote derivatives with respect to $\rho$. The
integration of the Einstein equations is simplified by imposing
the gauge condition \be\lb{gauge}  A = B +(\td+1)C + pD. \ee Using
the expressions for the Ricci tensor and substituting the form
field (\ref{solF}), we then find three equations for $B$, $C$ and
$D$ with similar differential operators \bea B'' &=& \frac{\td
b^2}{2(D-2)}\e^G, \label{EqB}\\ C'' &=&
-\frac{db^2}{2(D-2)}\e^G+ \td {\rm e}^{2(A-C)}, \lb{EqC}\\
D'' &=& \frac{\td b^2}{2(D-2)}\e^G, \label{EqD}  \eea where
\be\lb{G} G=a\phi+2B+2(d-1)D, \ee and the following constraint
equation \bea\lb{cons} & & -(B'+kC'+pD')^2 + B'^2 + kC'^2 + pD'^2
+ \frac12 \phi'^2 = \nonumber \\ & & \qquad\qquad =
\frac{b^2}2\e^G -  \td(\td+1)\e^{2(A-C)}.  \eea The dilaton
equation \be\frac1{\sqrt{-g}}\, \partial_\mu \left( \sqrt{-g}
\partial^\mu \phi \right) = \frac{a}{2\,(\td+1)!} {\rm e}^{a\phi}
F_{[\td+1]}^2  \label{dileq}\ee
 takes the following form
\begin{equation}\label{EqPhi}
\phi'' = \frac{a b^2}2\e^G.
\end{equation}

From the field equations it is clear that the functions $\B,\,\D$
and $\td\phi/(a(D-2))$ may differ only by a solution of the
homogeneous equation, which is a linear function of $\rho$. Thus
we have \bea
\D&=&\B+d_1\rho+d_0,\lb{DB}\\\phi&=&\frac{a(D-2)}{\td}\B+
\phi_1\rho+\phi_0,\lb{FB} \eea  where
$d_0,\,d_1,\,\phi_0,\,\phi_1$ are free constant parameters.
Substituting this into (\ref{G}) one finds  \be \lb{GB}
G=\frac{\de (D-2)}{\td}B+g_1\rho+g_0, \ee  with  \be
g_{0,1}=a\phi_{0,1}+2(d-1)d_{0,1},  \ee  so that the Eq.
(\ref{EqB}) becomes a decoupled equation for $G$: \be
G''=\frac{b^2\de }2\e^G. \ee  Its general solution, depending on
two integration constants $\alpha$ (real or imaginary) and
$\rho_0$ reads:  \be\lb{SolG} G=\ln\left(\frac{\alpha^2}{\de
b^2}\right) -\ln\left[\sinh^2\left(\frac{\alpha}{2}
(\rho-\rho_0)\right)\right], \ee  with $\alpha^2$ being the first
integral  \be\lb{al} G'^2-b^2\de\e^G=\alpha^2.  \ee  Combining the
Eqs. (\ref{EqB}),(\ref{EqC}) and (\ref{EqD}), one obtains for  the
linear combination  \be\lb{H}  H=2(A-C) = 2\td C + 2B + 2(d-1)D
\ee the second decoupled Liouville equation  \be\lb{EqH} H''=2
\td^2\e^H. \ee  The general solution depending on two parameters
$\beta,\,\rho_1$ reads  \be \lb{SolH}  H =
2\ln\beta/2\td-\ln\left( \sinh^2[\beta (\rho-\rho_1)/2 ] \right).
\ee  the first integral being \be \lb{Hint}  H'^2 -4 \td^2\e^H =
\beta^2.  \ee Finally, expressing the metric functions $\A,\,\C$
from (\ref{gauge}),(\ref{H}), one can write the full solution in
terms of $G,\,H$ as follows:  \bea
\B&=&\frac{\td}{\de(D-2)}\left(G-g_1\rho-g_0\right),\lb{Sol1} \\
\D&=&\frac{\td}{\de(D-2)}\left(G-g_1\rho-g_0\right)+d_1\rho+d_0,\lb{Sol2}
\\ \C&=&\frac{1}{2\td}\;H-\frac{d}{\de (D-2)}\;
G+c_1\rho+c_0,\lb{Sol3} \\ \A&=&
\frac{(1+\td)}{2\td}\,H-\frac{d}{\de(D-2)}\;G+c_1\rho+c_0,\lb{Sol5}\\
\phi&=&\frac{a}{\de}\;G+f_1\rho+f_0,\eea where\bea
c_{0,1}&=&\frac{a}{\de}\left(\frac{d}{D-2}\;\phi_{0,1}-
\frac{(d-1)a}{\td}\;d_{0,1}\right), \nonumber\\
f_{0,1}&=&\phi_{0,1}-\frac{a }{\de}\;g_{0,1} =
\frac{2\td}{a}c_{0,1}.\lb{Solc} \eea    Our solution depends on
nine parameters:
$b,\,d_0,\,d_1,\,\phi_0,\,\phi_1,\,\rho_0,\,\rho_1,\,\alpha,\,\beta$.
There remains to enforce the constraint following from Eq.
(\ref{cons})\bw \be\lb{cons1}
\frac{(\td+1)\beta^2}{4\td}-\frac{\alpha^2}{2\de}-
\frac{d\td}{\de(D-2)}\phi_1^2  + \frac{2a(d-1)}{\de}\phi_1d_1
\nonumber  -\frac{d-1}{\de(D-2)\td} \bigg[a^2(D-2)(D-3) +
2\td^2\bigg]d_1^2  = 0, \ee\ew  so that actually we have only eight
independent parameters.

\section{Solutions with regular event horizon}
The ansatz  (\ref{metric}) is invariant under  translations of $\rho$,
so that without  loss of generality we can choose  $\rho_1 = 0$. There
remain   seven  parameters.   Also,  the   results   (\ref{SolG})  and
(\ref{SolH}) do not  depend on the sign of $\rho$,  which we choose so
that the  horizon $\e^{2B} \to  0$ corresponds to $\rho  \to +\infty$,
with  $\alpha >  0$, $\beta  > 0$.   The main  divergent terms  in the
functions involved are \be G \sim -\alpha\rho, \quad H\sim -\beta\rho,
\ee        so       $B$        tends       to        \be       B\simeq
\frac{\td}{\Delta(D-2)}(-\alpha\rho-g_1\rho). \ee We are interested in
solutions   possessing  an   event  horizon,   that  is   a   zero  of
$$g_{tt}=\e^{2B}.$$  This  may  happen  while  $\rho\rightarrow\infty$
provided \be  \alpha+g_1>0.  \ee  In addition, we  have to  ensure the
regularity of  the horizon. From ()()  it follows that  $D$ and $\phi$
must  be finite  at the  horizon.  When  $\rho\to\infty$ we  have \bea
D\simeq  -\frac{\td}{\Delta(D-2)}(\alpha+g_1)\tau+d_1\tau\\ \phi\simeq
-\frac{a}{\Delta}\alpha\tau+f_1\tau,  \eea   so  the  coefficients  of
$\tau$ in $D$ and $\phi$ must vanish, which  gives the following two
relations       \bea       \lb{alp}       \alpha=\frac{\Delta}{a}f_1\\
d_1=\frac{\td}{a(D-2)}\phi_1.\lb{d1}  \eea  Thus  using the  shift
of $\rho$ and imposing the condition  of the regularity of the
horizon we have fixed  three parameters, and  five parameters
still  remain free. But we  can rescale $\rho$, $t$  et $x$
without  changing the physical meaning  of the  solution, this
allows  to fix  two more  parameters, namely, \be d_0=0,\qquad
d_1=1.\lb{d0} \ee Combining all the preceding conditions
(\ref{alp})-(\ref{d1}),   (\ref{d0})  and  the  constraint
(\ref{cons}),      we       obtain      \bea \alpha=\beta=2,\quad
f_{1}=\frac{2a}{\Delta},\quad        \phi_{1}=\frac{a}{\td},\nonumber\\
g_{1}=\frac{\Delta(D-2)}{\td}-2,\quad
c_{1}=\frac{a^2}{\Delta\td}, \eea so  only three  parameters
remain free:  $\rho_0$, $c_0$  et $b$. After  suitable rescaling
of  the brane  world-volume, the  resulting metric
reads            \bw\be            ds^2            =
\left(\frac{\e^\rho}{2\sinh(\rho-\rho_0)}\right)^{\frac{4\td}{\Delta(D-2)}}
\left(-\e^{-2\rho}dt^2+d{\bf
x}_p^2\right)
+\mu^2\left(\frac{\e^\rho}{2\sinh\rho}\right)^{\frac{2}{\td}}
\left(\frac{2\sinh(\rho-\rho_0)}{\e^\rho}\right)^{\frac{4d}{\Delta(D-2)}}
\left(d\Sigma^2+\frac{d\rho^2}{\td^2\sinh^2\rho}\right), \ee \ew
where $\mu$             is             defined            as
\be
\ln\mu=c_0+\frac{a^2}{\Delta\td}\ln2-\frac{1}{\td}\ln\td-\frac{d}{\Delta(D-2)}
\ln\left(\frac{4}{\Delta  b^2}\right). \ee  The  corresponding
dilaton function         is         \be
\e^{a\phi}=\frac{4\td^2}{\Delta
b^2}\mu^{2\td}\left(\frac{\e^\rho}{2\sinh(\rho-\rho_0)}\right)^
{\frac{2a^2}{\Delta}}.
\ee   Provided $\rho_0<0$, the  range of $\rho$ is  the positive
semi-axis,  with  $\rho  \to  +\infty$ corresponding  to  the
regular horizon, $\rho = 0$ to space-like infinity.

\section{$p$-branes with LDB asymptotics}
For   the  subsequent   analysis  it   is  convenient   to   make the
transformation  of  the  radial  coordinate \be  \lb{mu} e^{2\rho}  =
\frac{r}{r-\mu}, \ee with $r=\mu>0$  being the horizon radius, so
that \be       \sinh^2\rho      = \frac{\mu^2}{4r(r-\mu)},      \quad
\frac{d\rho^2}{\sinh^2\rho} = \frac{dr^2}{r(r-\mu)}.   \ee   Also,
putting \be e^{2\rho_0}  = \frac{r_*}{\mu-r_*}, \ee with $0 < r_* <
\mu$ (note that  $r_*$ is not an image of  $\rho_0$ with respect to
the map (\ref{mu})),  we obtain  for $g_{tt}$  the  following
expression:  \be g_{tt}=-\left(\frac{r_*(\mu-r_*)r^2}
{[(\mu-2r_*)r+r_*\mu]^2}\right)^{\frac{2\td}
{\Delta(D-2)}}\frac{r-\mu}{r}.   \ee An  examination  of this formula
shows that there  are two and only two  possibilities.  In the
generic case  $r_*  \neq \mu/2$  ($\rho_0  \neq  0$),  one obtains
the  usual asymptotically    flat    black    brane solutions
\cite{HoSt91, DuKhLu95,GaRy98}. In  the special case $r_* = \mu/2$
($\rho_0  = 0$), the  solution is  no longer asymptotically flat, and
reduces  to the two-parameter ($\mu$, $b$) configuration
\bea ds^2=\left(\frac{r}{\mu}\right)^{\frac{4\td}{\Delta(D-2)}}
\left(-\frac{r-\mu}{r}dt^2+d{\bf
y}_p^2\right)+\nonumber\\
\quad+\mu^2\left(\frac{r}{\mu}\right)^{\frac{2a^2}{\Delta\td}}
\left(d\Omega_{\td+1}^2+\frac{dr^2}{\td^2r(r-\mu)}\right)\lb{solnasD1}\\
\e^{a\phi}=\frac{4\td^2}{\Delta
b^2}\mu^{2\td}\left(\frac{r}{\mu}\right)^{\frac{2a^2}{\Delta}},\quad
F_{[\td+1]}=b  {\rm  vol}(\Omega_{\td+1}).\lb{nafphi}  \eea The
regular event horizon is at $r=\mu$, and if $\mu=0$ the coordinates
systems is not  well-behaved. To  improve  this  one has  to  rescale
the  radial variable, \be r=\frac{\mu}{c}\bar{r}^{\tilde{d}},\ee  to
introduce instead of $\mu$ a new parameter $c$ as follows
\be\mu=b^{\frac{2d} {\Delta (D-2)}}c^{\frac{a^2}{\Delta \tilde{d}}}
\ee and to rescale the coordinates $t$ and $x$ as follows \be
t\rightarrow\left(\frac{b}{c}\right)^{\frac{-2\tilde{d}}
{\Delta(D-2)}}t,\quad x\rightarrow\left(\frac{b}{c}\right)^
{\frac{-2\tilde{d}}{\Delta(D-2)}}x.   \ee In terms of the new
coordinates the solution reads after relabelling $\bar{r}\rightarrow
r$: \bea ds^2= \left(\frac{
r^\td}{b}\right)^{\frac{4\tilde{d}}{\Delta(D-2)}}
\left[-\left(1-\frac{c}{r^{\tilde{d}}}\right)dt^2+ d{\bf
y}_p^2\right] +\nonumber\\
+\left(\frac{b}{r^\td}\right)^{\frac{4d}{\Delta(D-2)}} \left[\left( 1
-\frac{c}{r^\td} \right)^{-1}dr^2 +r^2d\Omega^2_{\td+1}
\right],\nonumber\\ \e^{a\phi}=\frac{4\td^2}{\Delta}
\left(\frac{r^{\td}}{b}\right)^\frac{2a^2}{\Delta},\;\; F_{[\td+1]}=b
{\rm  vol}(\Omega_{\td+1}). \lb{brldb}\eea  This is the two-parameter
family of  asymptotically non-flat $p$-branes. When $r\to\infty$ the
solution approaches the linear  dilaton background (\ref{eldb}) with
the following    identification   of parameters    \be
r_0=b^{1/\td},\quad \e^{a\phi_{\infty}}=\frac{4\td^2}{\Delta}.
\lb{rel}   \ee

The quantity $\phi_{\infty}$ is no longer the asymptotic value of
the dilaton, but  rather  an ``inherited value'' from the
asymptotically flat $p$-brane (\ref{afsol}).  In terms of it the
dilaton function in (\ref{brldb})   reads   \be
\e^{(\phi-\phi_{\infty})}=
\left(\frac{r}{r_0}\right)^\frac{2a\td}{\Delta},  \ee  which
coincides with (\ref{dil}). The parameter $c$ measures the
strength of the singularity at $r=0$ and so is presumably
proportional to the mass of the black brane, as will be checked by
a quasilocal computation in Sect. 8 (Eq. (\ref{M2})). The
parameter $b$  from (\ref{brldb}) is proportional to the form
flux, or ``magnetic charge'' associated with the solution. It is
important to note \cite{ClGaLe02,ClLe04}, however, that this
charge is the same for all the members of the black brane family
$(b,c)$ living on the linear dilaton background $(b,0)$. Besides
this  charge associated with the background, it therefore makes
sense to define a ``proper brane charge'' as the difference
between the total charge of an asymptotically LDB  black brane
solution and that of the associated LDB vacuum. This proper charge
is identically zero in the present case (see also \cite{HaOb00}).

Near the horizon  the  space-time asymptotes to the product
$M_{p+2}\times S^{\td+1}$, similarly to the case
of the Schwarzschild geometry, this can  be easily  seen by
introducing the tortoise radial  coordinate. Thus  our  black
branes  on the  linear dilaton background interpolate between the
product of a flat space and a sphere near the horizon (with fixed
value of the dilaton) and the linear dilaton background at
infinity. This is somehow inverse to the situation with BPS
asymptotically   flat $p$-branes which interpolate between the
linear dilaton background at  the horizon and flat space (with
constant dilaton) at infinity.

Computing the scalar curvature for the solution (\ref{brldb}) one
gets \be\lb{R} R=\frac{4d\td^2}{\Delta^2(D-2)r^2}
\left(\frac{r^\td}{b}\right)^{\frac{4d}{\Delta(D-2)}}
\left(\Delta-\td-\frac{(D-2)a^2 c}{2d r^\td}\right).\ee At
infinity $r\to \infty$  \be R\sim r^{-\frac{2a^2}{\Delta}},\ee so
in the dilatonic case ($a\neq 0$) the scalar curvature vanishes.
In the non-dilatonic case $a=0$ one finds a constant value
throughout the whole space-time \be R=\frac{4
d\td^2(\Delta-\td)}{\Delta^2(D-2)}\ee independently of whether $c$
is zero or not (for $c=0$ the space-time coincides with
(\ref{dsdual}) and is the product $AdS_{p+2}\times S^{\td+1}$).
Note that for $2d=(D-2)$ (even $D$) and $a=0$ the lagrangian
possesses a conformal symmetry and thus $R=0$ identically.

Using the relations (\ref{rel}) one can pass to the dual frame via
the conformal transformation  (\ref{todual}): \bea \lb{brdual}
ds^2_{\rm dual}=\e^{\frac{2(2\td-\Delta)\lambda}{r_0\Delta}}
\left[-\left(1-\frac{c}{r^{\tilde{d}}}\right)dt^2+ d{\bf
y}_p^2\right] +\nonumber\\
\left(1-\frac{c}{r^\td}\right)^{-1}d\lambda^2+r_0^2
d\Omega^2_{\td+1}.\eea     This metric approaches the product
space  $AdS_{p+2}\times S^{\td+1}$ as $r\to \infty$, unless
$2\td=\Delta$. In this latter case the asymptotic space is the
product of the $p+2$ dimensional Minkowski space with a sphere,
$M_{p+2}\times S^{\td+1}$.

\subsection{NS5}
In the particular case $D=10, a=-1, p=5$ the solution (\ref{brldb})
reads \bea ds^2 &=& \left(\frac{ r}{r_0}\right)^{\frac12}
\left[-\left(1-\frac{c}{r^2}\right)dt^2+ d{\bf y}_5^2\right]
+\nonumber\\ &&+\left(\frac{r_0}{r}\right)^{\frac32}
\left[\left(1-\frac{c}{r^2}\right)^{-1}dr^2  +r^2d\Omega^2_3 \right],
\nonumber\\ \e^{-(\phi-\phi_\infty)} &=& \frac{r }{r_0},\qquad
F_{[3]}=r_0^2\,{\rm vol}(\Omega_3) \eea  ($r_0^2 = b$). Passing to
the string frame \be ds_{\rm
string}^2=\e^{(\phi-\phi_\infty)/2}ds^2,\ee and changing the radial
variable \be r=\sqrt{c}\cosh\sigma,\ee we obtain \bea ds_{\rm
string}^2 =-\tanh^2\sigma dt^2+r_0^2 d\sigma^2+d{\bf
y}^2_5+r_0^2d\Omega_3,\\ \e^{2(\phi-\phi_\infty)}=\frac{r_0^2}{c
\,\cosh \sigma},\quad F_{[3]}=r_0^2\,{\rm vol}(\Omega_3).\eea This is
the product of the two-dimensional black hole, a three-sphere and a
five-dimensional flat space, as found earlier in Ref. \cite{MaSt97}
and used as a holographic counterpart to the little string theory.

\subsection{Black holes}
Specifying   $d=1, p=0, \td=D-3$, we obtain the following
two-parametric family of multidimensional (magnetic) black holes
asymptotically approaching the linear dilaton background in the
Einstein frame \bw\bea ds^2= -\left(\frac{
r^{D-3}}{b}\right)^{\frac{4(D-3)}{\Delta(D-2)}}
 \left(1-\frac{c}{r^{D-3}}\right)dt^2
+\left(\frac{b}{r^{D-3}}\right)^{\frac{4}{\Delta(D-2)}}
\left[\left(1-\frac{c}{r^{D-3}}\right)^{-1} dr^2
+r^2d\Omega^2_{D-2}\right],\nonumber\\
\e^{a \phi }=\frac{4(D-3)^2}{\Delta}
\left(\frac{r^{D-3}}{b}\right)^\frac{2a^2}{\Delta},\;\;
F_{[D-2]}=b\, {\rm vol}(\Omega_{\D-2}). \lb{bhldb}\eea\ew
Transforming to the ``Schwarzschild`` radial coordinate \be
{\tilde r}= r^\nu \nu^{-1} b^{\frac{2(D-4)}{\Delta(D-2)}},\quad
\nu=1+\frac{2(D-4)(D-3)}{\Delta(D-2)},\ee one obtains the metric
\be ds^2=-U dt^2+\frac{d{\tilde r}^2}{U}+R^2 d\Omega_{D-3}^2,\ee
with \bea U=\left(\frac{
r^{D-3}}{b}\right)^{\frac{4(D-3)}{\Delta(D-2)}}
\left(1-\frac{c}{r^{D-3}}\right),\nonumber\\ R^2=r^2
\left(\frac{b}{r^{D-3}}\right)^{\frac{4}{\Delta(D-2)}},\eea which
was found earlier by Chan, Horne and Mann \cite{ChHoMa94} and
interpreted as ``charged dilaton black hole with unusual
asymptotic''.

In the non-dilatonic case $a=0$ one obtains \bea\label{nbba0} ds^2=
-\left(\frac{ r^{D-3}}{b}\right)^2
 \left(1-\frac{c}{r^{D-3}}\right)dt^2\nonumber\\
+b^{\frac{2}{(D-3)}} \left[\left(1-\frac{c}{r^{D-3}}\right)^{-1}
\frac{dr^2}{r^2} + d\Omega^2_{D-2}\right].\eea Contrary to
expectations, this is not a black hole. Changing the coordinates
${\hat r}=r^{D-3}-c/2$ one can rewrite this as the product space
$AdS_2\times S^{D-2}$. Thus the asymptotically LDB black holes exist
only in the dilatonic version. This is not true, however, for higher
$p$-branes.

\section{$p$-branes on $AdS_{p+2}\times S^{\td+1}$} For theories
without the dilaton, $a=0$, our general solution (\ref{brldb})
describes black branes on the background
$AdS_{p+2}\times S^{\td+1}$. The background is supported by the
antisymmetric form flux \be F_{[\td+1]}=r_0^\td {\rm
vol}(\Omega_{\td+1}).\ee In this case the Einstein and dual frames
coincide and we obtain \bea \lb{ndbrdual} ds^2
=\left(\frac{r}{r_0}\right)^{2\td/d}
\left[-\left(1-\frac{c}{r^{\tilde{d}}}\right)dt^2+ d{\bf
y}_p^2\right] + \nonumber\\ +\left(1-\frac{c}{r^\td}\right)^{-1}
\left(\frac{r_0}{r}\right)^{2}d r^2+r_0^2 d\Omega^2_{\td+1}.\eea
This metric describes a black $p$-brane on the
fluxed background $AdS_{p+2}\times S^{\td+1}$ for $p\geq 1$ (for
$p=0$ as we mentioned above the metric is gauge equivalent to the
background itself). The event horizon is at $r=c^{1/\td}$ and in
the near-horizon limit the geometry is the product space
$M_{p+2}\times S^{\td+1}$. Therefore the solution
(\ref{ndbrdual}) interpolates between the product space
$M_{p+2}\times S^{\td+1}$ at the horizon and $AdS_{p+2}\times
S^{\td+1}$ at infinity. Recall that the usual asymptotically flat
extremal non-dilatonic branes interpolate between $AdS_{p+2}\times
S^{\td+1}$ at the horizon (throat) and $M_D$  at infinity
\cite{DuGiTo94}.

Denoting our  brane solutions by tilde, we list the following
particular cases  \bw\bea \lb{ndbrdualm2} {\tilde{\rm M}2}:&&\quad
ds^2 =\left(\frac{r}{r_0}\right)^4
\left[-\left(1-\frac{c}{r^6}\right)dt^2+ d{\bf y}_2^2\right] +
\left(1-\frac{c}{r^6}\right)^{-1} \left(\frac{r_0}{r}\right)^{2}d
r^2+r_0^2 d\Omega^2_{7},\quad r_0=b^{1/6},\\{\tilde{\rm
M}5}:&&\quad ds^2 =\left(\frac{r}{r_0}\right)
\left[-\left(1-\frac{c}{r^3}\right)dt^2+ d{\bf y}_5^2\right] +
\left(1-\frac{c}{r^3}\right)^{-1} \left(\frac{r_0}{r}\right)^{2}d
r^2+r_0^2 d\Omega^2_{4},\quad r_0=b^{1/3},\\{\tilde{\rm
D}3}:&&\quad ds^2 =\left(\frac{r}{r_0}\right)^2
\left[-\left(1-\frac{c}{r^4}\right) dt^2+ d{\bf y}_3^2\right] +
\left(1-\frac{c}{r^4}\right)^{-1} \left(\frac{r_0}{r}\right)^{2}d
r^2+r_0^2 d\Omega^2_{5},\quad r_0=b^{1/4}.\eea \ew

These solutions are not supersymmetric unless $c=0$, in which case
they become $AdS_4\times S^7,\,AdS_7\times S^4$ and $AdS_5\times
S^5$ respectively.

\section{Quasilocal charges }%
In order  to find  the physical characteristics of  the
non-asymptotically flat solutions, one can use  the Hamiltonian
formulation of the problem similar  to that  applied earlier  to
four-dimensional  linear dilaton black holes \cite{ClGaLe02,ClLe04}.
Our approach closely follows that of Brown and  York \cite{BY}  and
Hawking and  Horowitz  \cite{HaHo95} (for  a recent  review  see
\cite{Bo00}).

The space-time metric is decomposed \`a la ADM \be\lb{ADM}
ds^2=-N^2dt^2+h_{ij}(dx^i+N^idt)(dx^j+N^jdt)  \ee where $N$  is
called the lapse function and $N^i$  the shift vector. This
decomposition means geometrically that the  space-time  is
foliated  by spacelike surfaces $\Sigma_t$, of metric
$h_{\mu\nu}=g_{\mu\nu}+u_\mu u_\nu$, labelled by a time coordinate
$t$ with the  unit normal  vector $u^\mu=-N\delta_0^\mu$. It
follows that the  timelike vector field $t^\mu$, satisfying
$t^\mu\nabla_\mu t=1$,  is decomposed into the lapse function and
the shift vector as $t^\mu=Nu^\mu+N^\mu$. The space-time boundary
$\p_M$ consists  of  initial and  final  surfaces $\Sigma_t,\,
t=t_i, t_f$  and a timelike surface $B$ to which the vector
$u^\mu$  is tangent.   This latter  surface is foliated by the
($D-2$)-dimensional   surfaces  $S^r_t$, of metric
$\sigma_{\mu\nu}=h_{\mu\nu}-n_\mu n_\nu$,   which   are
intersections   of $\Sigma_t$  and   $B$.   The   unit  spacelike
(outward)   normal  to $S^r_t$, $n^\mu$,  is orthogonal to
$u^\mu$.

In the following, we generalize the treatment of the four-dimensional
Einstein-Maxwell-dilaton  theory  \cite{Bo00} to the  case of  the
Einstein-dilaton-antisymmetric form theory in D dimensions.

The starting point is the $D$-dimensional
Einstein-dilaton-antisymmetric form action ({\ref{action})
supplemented by a boundary term \cite{RT} necessary to have a
well-defined variational principle
($q=\td+1$)\bea\label{actionemdDquas} &&S = \int_M d^D x \sqrt{-g}
\left( R_D - \frac12 \partial_\mu \phi
\partial^\mu \phi - \frac1{2\, q!} \, {\rm e}^{a\phi} \, F_{[q]}^2
\right)\nonumber\\&&+2\int_{\Sigma_{t_i}}^{\Sigma_{t_f}} K\sqrt{h}\,
d^{D-1}x-2\int_{B}\Theta\sqrt{\gamma}\, d^{D-1}x. \eea $K$ is the
trace of the extrinsic curvature $K^{\mu\nu}$ of $\Sigma_{t_{i,f}}$,
while $\Theta$ is the trace of the extrinsic curvature
$\Theta^{\mu\nu}$ of $B$, defined as \bea \lb{ec}
K_{\mu\nu}&=&-h_{\mu}^{\alpha}\nabla_\alpha u_\nu=
-\frac{1}{2N}\left(\dot{h}_{\mu\nu}-2D_{(\mu}N_{\nu)}\right),\\
\Theta_{\mu\nu}&=&-\gamma_{\mu}^{\alpha}\nabla_\alpha n_\nu,
\eea
where $\nabla$ and $D$ are the covariant derivatives compatible
with the metric $g_{\mu\nu}$ and $h_{ij}$, respectively.

The momentum variables conjugate to  $\phi$, $h_{ij}$ and
$A_{i_1\ldots i_{q-1}}$  are \bea p_\phi=-\sqrt{-g}\p^0\phi,\quad
p^{ij}=\sqrt h(h^{ij}K-K^{ij}),\lb{pij}\\ \Pi^{i_1\ldots
i_{q-1}}=-\sqrt{-g} \e^{a\phi}F^{0i_1\ldots i_{q-1}} \equiv -\sqrt
{-g} E^{i_1\ldots i_{q-1}}. \eea

First, consider the gravitational part of the action
(\ref{actionemdDquas}). To obtain the corresponding gravitational
contribution to the Hamiltonian, \be
H_G=\int_{\Sigma_t}\left(p^{ij}\dot{h}_{ij}-L_G\right)\sqrt{h}\,
d^{D-1}x, \ee we have used the well known result expressing the
$D$-dimensional scalar curvature in terms of the
$(D-1)$-dimensional scalar curvature, the extrinsic curvature of
$\Sigma_t$ and the total derivative terms \cite{HaHo95}, \bea
\lb{rd}
R_D=R_{D-1}-K^2+K_{\mu\nu}K^{\mu\nu}\nonumber\\+2\nabla_\mu(u^\mu\nabla_\nu
u^\nu)-2\nabla_\nu(u^\mu\nabla_\mu u^\nu), \eea and the following
relation \be \lb{pijhij}
p^{ij}\dot{h}_{ij}=-2N\sqrt{h}(K^2-K^{ij}K_{ij})+2p^{ij}D_{(i}N_{j)}.
\ee obtained by combining the relations (\ref{ec}) and (\ref{pij}).

Using these two results and the Eqs. (\ref{ADM}), (\ref{ec}), the
gravitational part of the Hamiltonian reads \bw \be
H_G=\int_{\Sigma_t}\sqrt{h}\left[N\left(-R_{D-1}+
K_{\mu\nu}K^{\mu\nu}-K^2\right)-N^iD_\mu \left(\frac{p^\mu{}_i}{\sqrt
h}\right)\right] +2\int_{S^r_t}\sqrt{\sigma}\left(Nk+\frac{n_\mu
p^{\mu\nu}N_\nu}{\sqrt h}\right)d^{D-2}x, \ee \ew where
$k=-\sigma^{\mu\alpha}D_{\alpha}n_\mu$ is the extrinsic curvature of
$S^r_t$ embedded in $\Sigma_t$. In the same way, the matter
contribution to the Hamiltonian is obtained using the following
relations \bw \bea
\partial_0\phi&=&\frac{N^2}{\sqrt{-g}}p_\phi+N^i\partial_i\phi,\quad
\partial^i\phi=h^{ij}\partial_j\phi+\frac{N^i}{\sqrt{-g}}p_\phi,\quad
F_{0i_2\cdots i_q}=N^jF_{ji_2\cdots i_q}+
\e^{-a\phi}\frac{N^2}{\sqrt{-g}}\Pi_{i_2\cdots i_q},\\
F^{i_1\cdots i_q}&=&\bar{F}^{i_1\cdots i_q}+\e^{-a\phi}
\left(N^{i_1}\frac{\Pi^{i_2\cdots i_q}}{\sqrt{-g}}-N^{i_2}
\frac{\Pi^{i_1i_3\cdots i_q}}{\sqrt{-g}}+\cdots-(-1)^qN^{i_q}
\frac{\Pi^{i_1\cdots i_{q-1}}}{\sqrt{-g}}\right), \eea \ew where
$\bar{F}$ is a $(D-1)$-dimensional tensor which indices are raised
and lowered by $h_{ij}$.

Finally, collecting the gravitational and matter parts, one
obtains the following expression for the Hamiltonian \bw\bea
H&=&\int_{\Sigma_t}\sqrt{h}\left(N{\cal H}+N^i{\cal H}_i+
A_{0i_2\ldots i_{q-1}}{\cal H}_A^{i_2\ldots
i_{q-1}}\right)d^{D-1}x
+2\int_{S^r_t}\sqrt{\sigma}\left(Nk+\frac{n_\mu
p^{\mu\nu}N_\nu}{\sqrt
h}\right)d^{D-2}x\nonumber\\
&&\quad +(q-1)\int_{S^r_t}N\sqrt{\sigma}A_{0i_2\ldots
i_{q-1}}E^{ji_2\ldots i_{q-1}}n_jd^{D-2}x, \eea where the
constraints  read \bea {\cal
H}=-R_{D-1}+K_{\mu\nu}K^{\mu\nu}-K^2+\frac{p_\phi^2}{2\sqrt
h}+\frac{\sqrt h}{2}(\p\phi)^2+\frac{\e^{-a\phi}}{2(q-1)!\sqrt
h}\Pi^2+\frac{\sqrt{h}}{2q!}\e^{a\phi}F^2,\\
{\cal H}_j=-D_\mu\left(\frac{p^\mu{}_j}{\sqrt
h}\right)+p_\phi\p_j\phi+\frac{1}{(q-1)!}\Pi^{i_1\ldots i_{q-1}}
F_{ji_1\ldots i_{q-1}},\\
{\cal H}_A^{i_2\ldots i_{q-1}}=-(q-1)\p_j\Pi^{ji_2\ldots i_{q-1}}.
\eea\ew

The quasilocal energy is defined as the ``on-shell'' value of the
Hamiltonian. Since the volume terms in the Hamiltonian are
proportional to constraints which vanish  for a solution of the
theory, the quasilocal energy of a solution is simply given by the
surface terms in the Hamiltonian, i.e. by quantities evaluated on the
$2$-surface $S^r_t$. We define the quasilocal mass as the quasilocal
energy evaluated in the limit $r\to\infty$. However, it is known that
the quasilocal energy is generally divergent at infinity. This
divergence may be regularized by subtracting the contribution of a
background solution, provided  one can impose the same Dirichlet
boundary conditions on $S^r_t$ for the solution under consideration
and for the background solution. Finally, the quasilocal angular
momentum of a solution may be obtained by carrying out an
infinitesimal gauge transformation $\delta\varphi=\delta\Omega \,t$
and evaluating the response  \be J=\frac{\delta H}{\delta \Omega}.
\ee

The resulting quasilocal energy and quasilocal angular momentum
are given by \bw \bea
E&=&2\int_{S^r_t}\sqrt{\sigma}\left(N(k-k_0)+\frac{n_\mu
p^{\mu\nu}N_\nu}{\sqrt h}\right)d^{D-2}x \nonumber\\
   &+& (q-1)\int_{S^r_t}A_{0i_2\ldots i_{q-1}}
(\bar{\Pi}^{ji_2\ldots
i_{q-1}}-\bar{\Pi}^{ji_2\ldots i_{q-1}}_0)n_jd^{D-2}x,\lb{eD}\\
J_i&=&-2\int_{S^r_t}\frac{n_\mu p^{\mu}{}_i}{\sqrt h}
\sqrt{\sigma}d^{D-2}x
 -(q-1)\int_{S^r_t}A_{ii_2\ldots i_{q-1}}
\bar{\Pi}^{ji_2\ldots i_{q-1}}n_jd^{D-2}x. \eea \ew  where
$\bar{\Pi}^{ji_2\ldots i_{q-1}}=(\sqrt \sigma/\sqrt
h)\Pi^{ji_2\ldots i_{q-1}}$. The quantities with the subscript $0$
are those associated with the background solution. Here, we have
written the formulas of the quasilocal quantities for a static
background solution ($N^i_0=0$ and $J_0=0$). Notice that the
dilaton does not contribute directly to the quasilocal energy and
quasilocal angular momentum.

Now, we are able to compute the mass and the angular momentum of
the solution (\ref{brldb}). Since the solution is static, the
angular momentum density $N^i=0$ is zero. Also, for our purely
magnetic solution $A_{0i_2\ldots i_{q-1}}=0$. So, only the first
term in (\ref{eD}) contributes to the quasilocal mass. The
extrinsic curvature of $S^r_t$ (a section $t=r=$ constant of
(\ref{brldb}) reads \be k=-\left(\frac{(1+\td)a^2}{\Delta}+
\frac{2\td^2(d-1)}{\Delta(D-2)}\right) b^{\frac{-2d}{\Delta(D-2)}}
r^{-\frac{a^2}{\Delta}}\sqrt{1-\frac{c}{r^\td}}, \ee so that
\be\lb{snk} \sqrt{\sigma}Nk=-\left(\frac{(1+\td)a^2}{\Delta}+
\frac{2\td^2(d-1)}{\Delta(D-2)}\right)
r^\td\sqrt{1-\frac{c}{r^\td}}. \ee At infinity
($r\rightarrow\infty$), the leading term in (\ref{snk}) diverges
($\td>0$) whereas the subleading term is finite. To obtain a
finite result, we subtract the contribution of a background
solution. The natural candidate is the vacuum of the black branes,
i.e. the LDB background obtained by taking the parameter $c$ equal
to zero. Subtracting this contribution cancels the leading term
and we get
 \be \lb{M2}\frac{{\cal
M}}{\mbox{vol}(\mbox{$p$-brane})}=\left(\frac{(1+\tilde{d})a^2}
{\Delta}+\frac{2\tilde{d}^2(d-1)}{\Delta(D-2)}\right)c
\;\mbox{vol}(\Omega_{\td+1}). \ee This expression is always positive
in view of the inequalities $d\geq 1$, $\Delta>0$ and $D>2$. The
first term may be interpreted as the dilaton contribution to the
mass, and the second as the proper brane contribution. Note that for
non-dilatonic branes $a=0$, one has a non-zero mass density only for
$p\geq 1$; the fact that in this case the would-be black holes
($d=1$) are actually massless is due to the gauge equivalence,
previously pointed out, between the solution (\ref{nbba0}) with $c
\neq 0$ and the background $c=0$.

Obviously, the solution (\ref{brldb}) being static, the quasilocal
angular momentum is equal to zero.

Now we address the question of the thermodynamics of the LDB black
branes (\ref{brldb}). In the case of the standard asymptotically
flat magnetostatic black branes, the first law is \be\lb{fl}
d{\cal M}=T dS, \ee  where the temperature is given by the inverse
period of imaginary time (or equivalently by the horizon surface
gravity over $2\pi$) and the entropy by the quarter of the horizon
area in planckian units. A simple calculation gives for the
temperature and the entropy of (\ref{brldb}),   \bea\lb{T}
T&=&\frac{\td}{4\pi}b^{-\frac{2}{\Delta}}c^{\frac{2}{\Delta}-
\frac{1}{\td}},\\ \frac{S}{\mbox{vol}(\mbox{$p$-brane})}&=&4\pi
b^{\frac{2}{\Delta}}c^{-\frac{2}{\Delta}+
\frac{1}{\td}+1}\mbox{vol}(\Omega_{\td+1})\lb{S} \eea (recall that
we use the value of the Newton constant $G=1/16\pi$). Then, using
(\ref{M2}) and (\ref{T})-(\ref{S}), the left-hand side and the
right-hand side of Eq. (\ref{fl}) read \bw \bea \frac{d{\cal
M}}{\mbox{vol}(\mbox{$p$-brane})
\mbox{vol}(\Omega_{\td+1})}&=&\left(\frac{(1+\tilde{d})a^2}
{\Delta}+\frac{2\tilde{d}^2(d-1)}{\Delta(D-2)}\right)\;dc,\\
\frac{TdS}{\mbox{vol}(\mbox{$p$-brane})
\mbox{vol}(\Omega_{\td+1})}&=&\left(\frac{(1+\tilde{d})a^2}
{\Delta}+\frac{2\tilde{d}^2(d-1)}{\Delta(D-2)}\right)\;dc+
\frac{2\td}{\Delta}\frac{c}{b}db. \eea \ew So, we see that the
first law is satisfied only if the parameter $b$, related to the
magnetic charge of the solution according to (\ref{nafphi}, is not
varied. This is consistent with our observation that this charge
is associated not with a specific black brane, but rather with the
LDB background. Since $b$ is not a parameter of the black branes,
it should not be varied \cite{ClGaLe02,ClLe04}. Then, we conclude
that the asymptotically LDB black branes satisfy the first law of
thermodynamics.

\section{Kaluza-Klein interpretation and rotation}
Recently the near-horizon limit of near extremal rotating branes was
discussed by Harmark and Obers \cite{HaOb00}. The field
configurations they derived have to be regarded as rotating
counterparts of the static asymptotically LDB $p$-branes discussed
here. It would be interesting to give a constructive derivation of
these solutions. However it seems difficult to obtain them via the
direct integration of the Einstein equations. Also, owing to the lack
of supersymmetry, they can not be found via the Bogomolny equations.
Here we suggest a transparent Kaluza-Klein procedure which could be
used to perform this goal in the case of asymptotically LDB electric
$p$-branes.

Consider the $D$-dimensional action \be \label{eaction} S_D = \int
d^D x \sqrt{-g_D } \left( R_D - \frac12 (\partial\phi)^2 -
\frac{\e^{a\phi}}{2(p+2)!} \, F_{(p+2)}^2 \right). \ee and
specialize to the electric $p$-brane sector ($p = d-1$) \bea
ds_{(D)}^2 & = &
\e^{-2p\beta/(\td+1)}\,g_{\mu\nu}\,dx^{\mu}\,dx^{\nu} +
\e^{2\beta}\,d {\bf y}^2, \\
F_{(p+2)} & = & F_{\mu\nu}\,dx^{\mu}\wedge dx^{\nu}\wedge
    dy^1\wedge\cdots\wedge dy^p\,,
\eea where all fields depend only on the $x^{\mu}$ ($\mu =
0,\cdots,\td+2$). The dimensional reduction to $D-p = \td+3$
dimensions leads to the action \bea\label{redaction} S_{\td+3} =
\int d^{\td+3} x \sqrt{-g} \Big( R -
\frac{p(D-2)}{\td+1}(\partial\beta)^2 \nonumber\\- \frac12
(\partial\phi)^2 - \frac14\e^{a\phi - (2p\td/(\td+1))\beta} \,
F_{(2)}^2 \Big). \eea The two scalar fields are decoupled by \bea
\beta & = & \alpha^{-1}\bigg(-\frac{\td}{D-2}\varphi +
a\eta\bigg), \label{diag1}\\ \phi & = & \alpha^{-1}\bigg(a\varphi
+ \frac{2p\td}{\td+1}\eta\bigg), \label{diag2} \eea with \be
\alpha^2 = a^2 + \frac{2p\td^2}{(D-2)(\td+1)}, \ee leading to \bea
\label{diagaction} S_{\td+3} = \int d^{\td+3} x \sqrt{-g} \Big( R
- \frac{p(D-2)}{\td+1}(\partial\eta)^2 \nonumber\\  -\frac12
(\partial\varphi)^2 - \frac14\e^{\alpha\varphi} \, F_{(2)}^2
\Big). \eea For the value \be \alpha^2 = 2\frac{\td+2}{\td+1}
\Longleftrightarrow a^2 = 4 - \frac{2d\td}{D-2} \qquad (\Delta =
4)\,, \label{Delta4}\ee this action is the sum of the action for
the harmonic field $\eta$ in $\td+3$ dimensions and of the
Kaluza-Klein dimensional reduction of the $\td+4$ Einstein-Hilbert
action
\begin{equation}\label{ein}
S_{\td+4} = \int d^{\td+4} x \sqrt{-g_{(\td+4)}}R_{(\td+4)}.
\end{equation}

This leads to a simple procedure to construct rotating asymptotically
LDB electric $p$-brane solutions of the action (\ref{action}) for the
values of $a^2$ satisfying (\ref{Delta4}) : 1) Start from the trivial
embedding in $\td+4$ dimensions of the $\td+3=D-p$ Myers-Perry
solution \cite{Myers}; 2) carry out a twisted dimensional reduction
to $\td+3$ dimensions; 3) oxidize the resulting solution to $D$
dimensions, taking into account an arbitrary harmonic function $\eta$
in (\ref{diag1})-(\ref{diag2}). The associated rotating magnetic
$p$-brane solution may be derived from this by the electric-magnetic
duality transformation (\ref{duality}). In the case $p= D - 4$,
rotating asymptotically LDB dyonic $p$-branes have also been
generated \cite{Le04} by a procedure generalizing that used for $D =
4$ in \cite{ClLe04}.

Let us check that, in the static case, this procedure reproduces the
asymptotically LDB electrostatic $p$-brane already found. The
$\td+3$-dimensional Tangherlini metric embedded (with a twist $dt \to
dt - d\chi$) in $\td+4$ dimensions is \bw\bea ds_{\td+4}^2 & = &
d\chi^2 - \bigg(1-\frac{\mu}{\rho^{\td}}\bigg)(dt-d\chi)^2 +
  \bigg(1-\frac{\mu}{\rho^{\td}}\bigg)^{-1}\,d\rho^2 +
    \rho^2\,d\Omega_{\td+1}^2 \nonumber \\
& = & d\chi^2 - \bigg(1-\frac{\mu}r\bigg)(dt-d\chi)^2 +
    r^{2/\td}\bigg(\frac1{\td^2}\frac{dr^2}{r(r-\mu)} +
    d\Omega_{\td+1}^2\bigg) \nonumber \\
& = &
\e^{\frac{\td+1}{\td+2}\alpha\varphi}\bigg(d\chi+A_0\,dt\bigg)^2 +
\e^{\frac{-1}{\td+2}\alpha\varphi}\,ds_{\td+3}^2\,, \eea with \bea
\e^{\varphi/\alpha} & = & \bigg(\frac{\mu}r\bigg)^{1/2}\,, \quad
A_0 =
- \frac{r-\mu}{\mu}\,, \\
ds_{\td+3}^2 & = &
-\bigg(\frac{r}{\mu}\bigg)^{\frac{\td}{\td+1}}\frac{r-\mu}r\,dt^2
+
  \mu^{\frac2{\td}}\bigg(\frac{r}{\mu}\bigg)^{\frac{\td+2}{\td(\td+1)}}
\bigg(\frac1{\td^2}\frac{dr^2}{r(r-\mu)} +
    d\Omega_{\td+1}^2\bigg)
\eea \ew (the twisted dimensional reduction also involves a scale
$r_0$ which we have omitted). Carrying out the third step
(oxidization) then leads, for the choice $\eta=0$, to \bea ds_{(D)}^2
&=&
\bigg(\frac{r}{\mu}\bigg)^{\frac{\td}{D-2}}\bigg(-\frac{r-\mu}r\,dt^2
+ d{\bf y}^2\bigg) \nonumber \\   &+&
  \mu^{\frac2{\td}}\bigg(\frac{r}{\mu}\bigg)^{\frac{a^2}{2\td}}
\bigg(\frac1{\td^2}\frac{dr^2}{r(r-\mu)} +
    d\Omega_{\td+1}^2\bigg)\,, \\
\e^{a\phi} & = & \bigg(\frac{r}{\mu}\bigg)^{\frac{\td}{D-2}}\,.
\eea This is essentially the asymptotically LDB black $p$-brane
\footnote{For the reduced ($\td+3$)-metric above, the generic
spherically symmetric harmonic function $\eta$ is $\eta =
c\,\ln((r-\mu)/r) + d$.  It follows that for the generic ($\eta
\neq$ constant) asymptotically LDB $p$-brane solution, the dilaton
$\phi$ and the $p$-brane metric function $\beta$ are singular on
the horizon.}.

\section{Uniqueness, cosmic censorship and supersymmetry}
Combining the results of the previous analysis we can formulate the
following uniqueness theorem for static $p$-branes with a spherical
transverse space:
\begin{quote}
{\bf Theorem:} {\em A singly charged $p$-brane solution of the
Einstein equations with the dilaton and antisymmetric form sources
possessing a regular event horizon and the $R\times ISO(p)\times
SO(D-p-1)$ isometries is either asymptotically flat, in which case it
coincides with the standard black brane solution, or asymptotically
LDB and then it is given by Eq. (\ref{brldb})}.
\end{quote}

These two families are geometrically ``dual'' in the following sense.
The standard asymptotically flat dilatonic $p$-branes possess the BPS
limit in which the solutions have a null singularity. The BPS
dilatonic branes interpolate between the LDB at the horizon and
Minkowski space with a constant dilaton at infinity. The second
family interpolates between the product of flat space with a sphere
(with fixed dilaton) at the horizon and LDB at infinity. For the
second family the BPS limit coincides with the LDB itself.

Our proof was based on a complete integration of the corresponding
system of equations and subsequent determination of the free
parameters from physical requirements. It turns out that if one
imposes first the condition of existence of a regular event
horizon, there remain two and only two options for the asymptotic
behavior: either the solution is asymptotically flat, or it is
asymptotically LDB. A non-trivial feature of this situation is
that both asymptotic configurations are supersymmetric in ten and
eleven-dimensional supergravities and their toroidal dimensional
reductions. Therefore, demanding that the curvature singularity be
hidden behind the event horizon, i.e. imposing the cosmic
censorship requirement, we find that  non-supersymmetric black
brane solutions always ``choose'' a supersymmetric asymptotic.
This ``choice'' of supersymmetric asymptotics by the solution
satisfying a cosmic censorship, is another aspect of the
relationship which was called in Ref. \cite{KA} ``supersymmetry as
a cosmic censor''. Though this has been proven here under the
assumption of $R\times ISO(p)\times SO(D-p-1)$ isometries (i.e.
for singly charged branes), we expect this to be also true for
intersecting branes \cite{int} (for other early references see
\cite{Ga97}, intersecting solutions with extra parameters were
also  found recently \cite{MiOh04}).

It is worth noting that various attempts \cite{nonbps} were made
recently to interpret the  supergravity $p$-brane solution
(obtained under the same ansatz as here) with other free
parameters than the mass, the charge and the asymptotic value of
the dilaton as describing non-BPS string theory branes (fractional
branes, brane-antibrane systems ect.) Some of these solutions were
used in a formal way without careful checking of regularity and
asymptotic behavior. In view of the results of Ref.
\cite{GaLeCl04}, all solutions with extra parameters other than
the mass, the form charge and the asymptotic value of the dilaton
are either asymptotically non-flat or contain naked singularities.
Here we have proved in addition that the two-parameter family of
asymptotically non-flat $p$-branes without naked singularities is
necessarily a $p$-brane with LDB asymptotics. We realize, however,
that if cosmic censorship is not imposed (which may have some
justification in the string theory context) the solutions with
extra parameters can be useful.

No uniqueness of this kind is expected for the stationary
solutions. Indeed, as was extensively discussed recently, the
higher-dimensional Kerr solution is by no means unique within the
class of regular asymptotically flat stationary solutions, and
other asymptotically flat solutions such as rotating rings
\cite{EmRe02} and their generalizations (for recent references see
\cite{HoRe04}) including supersymmetric  configurations
\cite{El04} exist with horizon topologies other than a sphere.
Whether these solutions admit asymptotically non-flat counterparts
as in the case of the asymptotically LDB branes discussed here
remains an open question.

\section{Conclusions}
In this paper we have constructively derived the static black
$p$-brane solutions to supergravity theories which asymptotically
approach the linear dilaton background . The latter is the bulk
configuration which was interpreted as a holographic dual to
certain non-conformal quantum field theories (DW/QFT
correspondence) or (in the case of the NS5 brane) to little string
theory.

In the static case the LDB space-time in the dual frame factorizes
into the product of an AdS space and a sphere. There are two
substantially different cases: dilatonic, in which the dilaton is
linear in terms of a special radial variable, and non-dilatonic, when
the asymptotic space-time has the above factorization  property also
in the Einstein frame. This second case is fully supersymmetric,
while in the first one the supersymmetry is partially broken by the
dilaton. $p$-brane solutions asymptotically approaching the LDB
space-time exist only in the black version and are not
supersymmetric; their BPS limit coincides with the linear dilaton
background  itself. In the non-dilatonic case the lower member of the
family $p=0$ (black hole) does not exist, while in presence of the
dilaton all $p$ are possible (though we didn't study explicitly the
domain wall and instanton cases).

We have presented the generalization of the
Brown-York-Hawking-Horowitz formalism of quasilocal charges to the
case of arbitrary dimensions and the presence of antisymmetric form
fields. Using it we have shown that the asymptotically LDB $p$-branes
satisfy the first law of thermodynamics. In this derivation a finite
value for the asymptotic mass was obtained by subtracting the
infinite contribution of the background LDB. It is worth noting that
the charge parameter associated with the solution should not be
varied, being a property of the background rather than of a specific
brane. Therefore one may consider the asymptotically LDB branes to be
essentially uncharged with respect to the form field (this nicely
fits with the fact that their BPS limit is the LDB itself).

Some stationary solutions of the type discussed here were obtained
previously as near-horizon limits of near-extremal spinning black
branes, and here we have suggested  a constructive procedure for
their derivations as solutions to the supergravity field
equations. This also opens an interesting question about the
possible existence of asymptotically non-flat rotating rings. A
more obvious possible generalization could be to intersecting
branes of the type considered here.

\begin{acknowledgments}
The authors would like to thank Chiang-Mei Chen for useful
discussions and Michael Duff for historical comments. D.G. is
grateful to LAPTH Annecy for hospitality in November 2004 while
the paper was written. C.L. wishes to thank GReCO/IAP for
hospitality when this work was finalized. The work of D.G. was
supported in part by the RFBR grant 02-04-16949.
\end{acknowledgments}

\end{document}